\documentclass[twocolumn]{aastex701}

\usepackage{amsmath}
\usepackage{mathtools}
\usepackage[OT1]{fontenc}

\newcommand{\Ct}{$^{12}${\rm{C} }}

\newcommand{\Bt}{$^{10}${\rm{B}}}

\newcommand{\Bel}{$^{11}${\rm{B} }}

\newcommand{\pc}{{\mathrm{pc}}}

\newcommand{\Myr}{{\mathrm{Myr}}}

\newcommand{\eV}{{\mathrm{eV}}} 

\newcommand{\cm}{\mathrm{cm}} 
\newcommand{\km}{\mathrm{km}} 
\newcommand{\s}{\mathrm{s}} 
\newcommand{\mbarn}{\mathrm{mbarn}} 

\newcommand{\antoine}[1]{}

\usepackage{todonotes}

\begin{document}

\title{Magnetic Reconnection Modifies Cosmic-Ray Transport in the Interstellar Medium}

\author[0000-0002-5376-3322]{Antoine Baldacchino-Jordan}
\affiliation{Institute of Astronomy, Faculty of Physics, Astronomy and Informatics, Nicolaus Copernicus University in Toruń,
ul. Grudziadzka 5/7, PL-87-100 Toruń, Poland}
\email{503332@doktorant.umk.pl}
\author[0000-0002-2370-5631]{Michal Hanasz}
\affiliation{Institute of Astronomy, Faculty of Physics, Astronomy and Informatics, Nicolaus Copernicus University in Toruń,
ul. Grudziadzka 5/7, PL-87-100 Toruń, Poland}
\email{mhanasz@umk.pl}
\author[0009-0008-0241-6591]{Vinod Pisharody}
\affiliation{Leibniz Institute for Astrophysics, An d. Sternwarte 16, 14482 Potsdam, Germany}
\email{mogrodnik@abs.umk.pl}
\author[0000-0003-3054-6461]{Dominik Wóltański}
\affiliation{Institute of Astronomy, Faculty of Physics, Astronomy and Informatics, Nicolaus Copernicus University in Toruń,
ul. Grudziadzka 5/7, PL-87-100 Toruń, Poland}
\email{wolt@umk.pl}
\author[0000-0002-4435-4594]{Artur Gawryszczak}
\affiliation{Nicolaus Copernicus Astronomical Center, Bartycka 18, PL-00-716 Warsaw, Poland}
\email{gawrysz@gmail.com}

\begin{abstract}

We present numerical magnetohydrodynamic (MHD) simulations of the interstellar medium in gravity-stratified boxes with the {\scshape Piernik} MHD code. The system contains thermal gas, spectrally resolved cosmic-ray (CR) nuclei within the CRESP algorithm, and spectrally unresolved CR protons. CR protons drive gas outflows and undergo Coulomb and hadronic losses. CR nuclei undergo Coulomb losses. We also include the spallation of primary Carbon by thermal gas atoms, thereby decaying Carbon into secondary stable Boron isotopes. We use the resistive-MHD approximation as a subgrid model for unresolved turbulent magnetic dissipation in current sheets. We perform a parameter study of the magnetic diffusivity $\eta$. The Boron-to-Carbon ratio B/C and hadronic $\gamma$-ray luminosity density are computed as diagnostics. We find that non-zero magnetic diffusivity modifies the magnetic topology, leading to a more horizontal field in the halo, thereby accumulating CRs in the disk region and significantly increasing the CR escape time. 
As a consequence, the local turbulent conditions and turbulent reconnection in the interstellar medium should be considered when interpreting B/C and hadronic $\gamma$-ray luminosity observed at Earth. Our results suggest that the escape time and diffusion coefficient of CRs deduced from observational data in the vicinity of the Solar System cannot be extrapolated to the whole Galaxy.

\end{abstract}

\keywords{\uat{Cosmic Rays}{573} --- \uat{Magnetohydrodynamics}{343} --- \uat{High Energy astrophysics}{739} --- \uat{Interstellar medium}{847}}

\section{Introduction}

Cosmic rays (CRs), a non-thermal population of charged subatomic particles traveling within galaxies and galaxy clusters, are an essential component of the interstellar medium (ISM) \citep{2007arnps..57..285s,2015ara&a..53..199g}. CRs carry a total kinetic energy density comparable to the turbulence of the ISM gas and magnetic-field energy density in the Galaxy \citep{2015A&ARv..24....4B}, making them dynamically important for the ISM. Numerical simulations of the Local ISM and galactic disk have shown that CRs can generate large-scale gas outflows \citep{2013ApJ...777L..38H,2022MNRAS.510.3917G,2024MNRAS.52710897G,2024ApJ...964...99A, 2025A&A...700A.124K}. This process, named CR feedback, is an active topic of research \citep{2023A&ARv..31....4R}. Meanwhile, understanding CR propagation in the magnetized ISM and the galactic medium remains a major problem in CR physics \citep{2007arnps..57..285s}.

Today, several CR observables are accessible. CR electrons produce inverse Compton scattering emission by interacting with photons from the ISM and the cosmic microwave background, and relativistic electrons accelerated in magnetic fields generate synchrotron emission. Those emissions can be detected using multiwavelength observations \citep{10.1111/j.1365-2966.2011.19114.x, 2011A&A...534A..54S}. On the other hand, CR protons produce secondary pions from inelastic collisions against the thermal gas atomic nuclei, emitting detectable $\gamma$-rays that provide constraints on CR transport and feedback \citep{2009Natur.462..770V, 2018A&A...617A..73H, 2022ApJS..260...53A}. Moreover, collisions of primary CRs, such as Carbon, Nitrogen, and Oxygen, with ISM gas produce secondary particles, including Boron and Beryllium. The Boron-to-Carbon ratio B/C, detected, for example, by the AMS-02 spectrometer in the International Space Station \citep{2016phrvl.117w1102a, 2021PhR...894....1A}, provides an important constraint on the diffusion coefficient of particles \citep{2019phrvd..99j3023e}. Altogether, those observables provide key data about CR transport and interaction parameters. However, the degeneracy among possible transport modes (advection with gas, streaming with self-excited Alfvén waves, and diffusion within randomly perturbed magnetic fields) and collisional processes makes determining CR transport properties non-trivial.

The interstellar magnetic field plays a key role in the dynamics of spiral galaxies \citep{2004Ap&SS.289..293B}. Observations of large-scale polarized emission have revealed regions in spiral galaxies with strong, regular magnetic fields between the spiral arms, commonly referred to as magnetic arms \citep{2007A&A...470..539B, 2019Galax...8....4B}. Recent studies have found that the hot gas is spatially correlated with the inter-arm regions, suggesting that some gas heating process is physically related to the formation of these magnetic arms and proposing that magnetic reconnection is a plausible mechanism for heating the interstellar gas \citep{2016A&A...585A...3W, 2020A&A...640A.109W, 2022A&A...665A..64W, 2022A&A...664A.108W}.

The magnetic reconnection process involves the formation of tangential discontinuities in magnetic fields associated with thin current sheets, where ohmic dissipation converts antiparallel components of the magnetic field on both sides of the current sheet, thereby reconnecting previously disconnected field lines. 
This process can potentially change the global magnetic configuration of an astrophysical object. A modern view of magnetic reconnection under astrophysical conditions points towards turbulent reconnection \citep[see theoretical details in][]{Lazarian_1999,2020PhPl...27a2305L}, in which the magnetic field dissipation and transformation of magnetic topology occur at many microscopic reconnection sites rather than in large-scale current sheets.

In spiral galaxies, star formation is primarily localized in the spiral arms, with strong turbulence driving in those regions and therefore a dominance of turbulent magnetic fields over regular fields. The interarm regions are less affected by the star-formation activity within spiral arms, and the regular, predominantly horizontal component of the magnetic field dominates here. Examples of observations supporting this picture are  NGC 6946, M51, M74, and other galaxies \citep{2019Galax...8....4B}, with magnetic arms situated between the material arms. The canonical model of the spiral structure \citep{binney2008galactic} predicts that gas and stars flow through the spiral arms, with velocity corresponding to the difference between the pattern speed and the circular velocity at a given radius. The magnetized turbulence, strongly enhanced by star formation in spiral arms, moves away from the dense arms together with the fluid flow. Faraday depolarization analysis indicates that the magnetized ISM is less turbulent in the interarm regions than in spiral arms \citep{2007A&A...470..539B}.  A natural process accompanying the flow of the turbulent ISM is the turbulent cascade, with turbulent reconnection serving as a key element of magnetic turbulence. In strong MHD turbulence, magnetic lines constantly reconnect at all scales, making reconnection an intrinsic component of the turbulent cascade \citep{Lazarian_1999,2020PhPl...27a2305L}. This scenario is supported by X-ray observations of some galaxies, with NGC 6946 as a prime example: the X-ray emission indicates a spatial correlation between the temperature of the hot gas and the inter-arm regions, which might be interpreted as a heating effect caused by magnetic reconnection \citep{2016A&A...585A...3W, 2020A&A...640A.109W, 2022A&A...665A..64W, 2022A&A...664A.108W}.

The presence of magnetic reconnection from localized turbulent magnetic fields in spiral galaxies may have implications for galactic CR transport. Recent studies of CR feedback in galaxies incorporate the concept of anisotropic, magnetic-field-aligned CR propagation, including diffusion alone or in combination with streaming \citep{2020LRCA....6....1M,2023A&ARv..31....4R}. The magnetic field structure determines the propagation routes of CR particles and influences their residence time in the galactic disk \citep{2026ApJS..284...34B}. We therefore expect that the global magnetic-field configuration is as relevant as CR transport parameters in shaping the spatial distribution of observables such as $\gamma$-ray emission and B/C. Since charged CRs mostly propagate along magnetic field lines, magnetic reconnection may annihilate some of the vertical, wind-generated magnetic field and locally trap CRs within closed loops, increasing their escape time in the magnetic arms. Another possibility is that the reconnection can open closed field lines and eject CRs. If those scenarios are correct, there should be a difference in the CR transport parameter values between the arm and inter-arm regions. The CR escape time measured within our solar system may therefore not be interpreted as a global property of galactic CRs.

In this paper, we investigate the effect of topological evolution of galactic magnetic fields, driven by resistive processes, on the propagation of CRs and their observational characteristics. For this purpose, we employ self-consistent, CR-Magnetohydrodynamic (CR-MHD) simulations within the {\scshape Piernik} MHD code \citep{2010eas....42..275h, 2010eas....42..281h, 2012eas....56..363h, 2012eas....56..367h}. As in our previous paper \citep{2026ApJS..284...34B}, we intend to carry out simulations of the evolution of the ISM perturbed by CR injection in supernova remnants, while neglecting their thermal feedback. To this end, we use a model of a local patch of the galactic disk represented by a tall box, stratified by vertical gravity.

Despite limiting the domain size to achieve the highest possible simulation resolution, the high computational cost of our multi-component CR simulations for spectral description does not permit sufficient spatial resolution to model turbulent reconnection directly \citep{2011ApJ...735..102K, 2026hepr.confE..12K}. Therefore, we adopt a simple ohmic magnetic dissipation model as a subgrid realization of the turbulent reconnection model.
We assume that turbulent diffusivity of the ISM is of the order of  $\eta_{\rm turb}=L_{\rm turb} v_{\rm turb}/3 \sim 10^{25} - 10^{26} \cm^2\s^{-1}$, following from the characteristic scale of interstellar turbulence $L_{\rm turb} \sim 100 \pc$ and the characteristic turbulent velocity $v_{\rm turb} \sim 10 \km\s^{-1}$ \citep{2004astro.ph.11739S,Shukurov_Subramanian_2021}.

We perform a parameter study of the magnetic diffusivity $\eta$ with four simulations, including one ideal-MHD scenario ($\eta = 0$) for comparison. We analyze the impact of the resistive plasma on variations in the spatial distributions of B/C and hadronic $\gamma$-ray luminosity density. B/C as a function of kinetic energy per nucleon is obtained from spectrally-resolved primary and secondary CRs using the CRESP algorithm \citep{2021apjs..253...18o, 2026ApJS..284...34B}. The hadronic $\gamma$-ray luminosity density is retrieved in post-processing using the energy-loss rate of non-spectral CR proton collisions with thermal gas protons, thereby generating secondary neutral pions that decay into $\gamma$-rays.

The layout of this paper is as follows: in Section \ref{sec:simsetup}, we detail the simulation setup and describe the methodology for modeling CRs in our MHD simulations within {\scshape Piernik}. Sect. \ref{subsec:results} presents the results and discusses how magnetic diffusivity affects ISM dynamics and CR transport. In Sect. \ref{subsec:conclusion}, we discuss the results and conclude this paper.

\section{Simulation setup} \label{sec:simsetup}

We perform a series of gravity-stratified box simulations using the {\scshape Piernik} MHD code to represent the ISM in the neighborhood of the Solar System. The setup is the same as that presented in \cite{2026ApJS..284...34B}, but has four notable differences:
\begin{enumerate}
\item The numerical resolution is increased by a factor of two to reduce the numerical resistivity.
\item The non-spectral CR protons undergo Coulomb and hadronic losses.
\item Only \Ct, \Bt, and \Bel stable nuclei isotopes are present to limit the computational cost.
\item The magnetic field is solved in the resistive-MHD limit.
\end{enumerate}
In the following section, we detail this setup.

\subsection{Gravity-stratified box setup}

We consider a local rectangular patch of the ISM in initial hydrostatic equilibrium of dimension $L_x \times L_y \times  L_z = 0.5 \;\mathrm{kpc} \times 1 \;\mathrm{kpc} \times 8 \;\mathrm{kpc} $ in $x$, $y$, $z$ directions, with minimum cell size $\Delta x=\Delta y=\Delta z= 31.25\;\mathrm{pc}$. The box is stratified by vertical gravity. We choose physical conditions typical for the solar neighborhood in the Galaxy, with the thermal gas number density $ n_{\mathrm{ISM}}\approx 1 \;\mathrm{cm}^{-3}$ and temperature $T \approx 7000 \;\mathrm{K}$ at the galactic mid-plane \citep{1998ApJ...497..759F}. We assume an initial magnetic field $B_y = 3 \;\mathrm{\mu G}$ along the $y$ axis. We choose periodic boundary conditions for the $x$-axis and $y$-axis, and outflow boundary conditions in the $z$-direction. 
For the diffusing cosmic ray components, we set the boundary values of the number and energy densities to the floor values.

We apply the ohmic resistivity model together with the resistive MHD system of equations presented in Appendix \ref{Appendix_A},  as a subgrid model for turbulent reconnection.   The resistive-MHD solver is presented in detail in Appendix \ref{Appendix_B}.
In our numerical model, magnetic field dissipation occurs in large-scale current sheets, whereas in a real turbulent ISM, it is expected to occur in small-scale current sheets of turbulent flows.

Having defined the initial equilibrium state, we inject primary CRs in randomly-located supernovae \citep{2004ApJ...605L..33H}. Supernova events are continuously injected at a rate of $80\,\mathrm{kpc}^{-2}\,\mathrm{Myr}^{-1}$. We assume that each supernova remnant provides CRs with 10\% of the supernova's energy $E_{\mathrm{SN}}=10^{51} \,\mathrm{erg}$ for the proton component. We dynamically inject spectrally unresolved proton energy density, coupled to the magnetized thermal gas of the ISM, via the CR pressure gradient that drives gas outflows. In this study, we only consider spectrally-resolved primary \Ct with relative abundance to proton $N_C/N_p = 4.5 \times 10^{-4}$. We assume, for simplicity, that no thermal or kinetic energy is input from supernovae. \Ct decays from spallation with the ISM gas, producing secondary \Bt\, and \Bel.

We model spectrally-resolved CR nuclei with spallation, adiabatic, and Coulomb losses within the CRESP algorithm in {\scshape Piernik}, following the method presented in \cite{2001cophc.141...17m, 2020mnras.491..993g, 2021apjs..253...18o}, and refer the reader to the previous paper \citep{2026ApJS..284...34B} for details of the numerical implementation.

\subsection{CR protons and $\gamma$-ray post-processing}

In all our simulations, we apply the one-bin (grey) approximation to represent the CR proton energy density $e_{\mathrm{p}^+}$. Integrating the CR propagation equation over the entire momentum space eliminates the explicit momentum dependence, yielding the system of equations (\ref{eq.1}) and (\ref{eq.2})  presented in the appendix \ref{Appendix_A}, following \cite{1985A&A...151..151S}.

In post-processing, we compute the $\gamma$-ray luminosity density from proton-proton collisions \citep{2011hea..book.....l}

\begin{eqnarray}
\label{eq.3}
\mathcal{L} &=& \frac{1}{3} n_\mathrm{ISM} \sigma^\pi_{pp} c e_{\mathrm{p}^+} \\
&=& \frac{\sigma^\pi_{pp} c}{3 m_p} \rho_\mathrm{ISM} e_{\mathrm{p}^+} \nonumber \\
&=& 1.2 \times 10^{-14} \left(\frac{\rho_\mathrm{ISM}}{M_\odot\,\pc^{-3}}\right) \left(\frac{e_{\mathrm{p}^+}}{\eV \,\cm^{-3}}\right)\,\eV \,\cm^{-3} \,\s^{-1}, \nonumber
\end{eqnarray}
where $m_p$ is the proton mass, $c$ the speed of light and $\sigma_{pp}^\pi \approx 32 \, \mbarn$ the cross section of the pion-producing proton-proton collision \citep{2004A&A...413...17P}. The $1/3$ term accounts only for neutral pion production.

\section{Results} \label{subsec:results}

\subsection{Impact of difusivity on the ISM dynamics}

\begin{table}
\centering
\begin{tabular}{llllllll}
\hline
Model    &  $\eta$ ($\mathrm{pc}^{2}\mathrm{Myr}^{-1}$) & $\eta$ $(\mathrm{cm}^2 \mathrm{s}^{-1})$ \\ \hline

A & $0$                                                                                     & $0$   \\
B  & $30$                                                                                     & $9.0\times 10^{24}$                                                                                                                                                        \\
C & $60$                                                                                     & $1.8\times 10^{25}$                                                                                                                                                            \\
D &  $100$                                                                                     & $3.0\times 10^{25}$   \\
\end{tabular}
\caption{The different simulation models (left) with different values for the magnetic diffusivity. The chosen values for $\eta$ are displayed in units of $\mathrm{pc}^{2}\mathrm{Myr}^{-1}$ (middle) and $\mathrm{cm}^2 \mathrm{s}^{-1}$ (right).}
\label{table_1}
\end{table}

\begin{figure*}[ht]
    \centerline{
    \includegraphics[width=.5\linewidth]{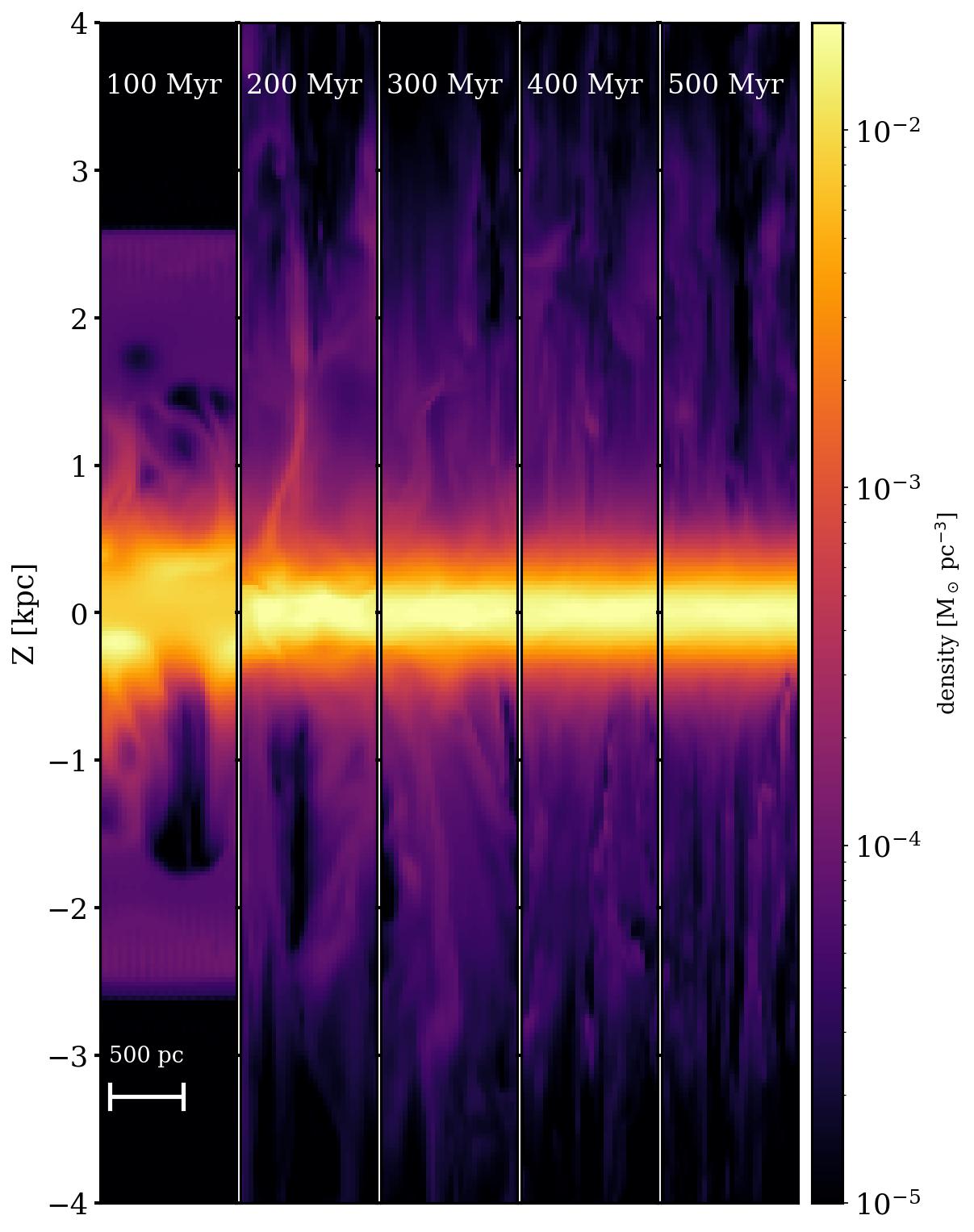
}
    \includegraphics[width=.5\linewidth]{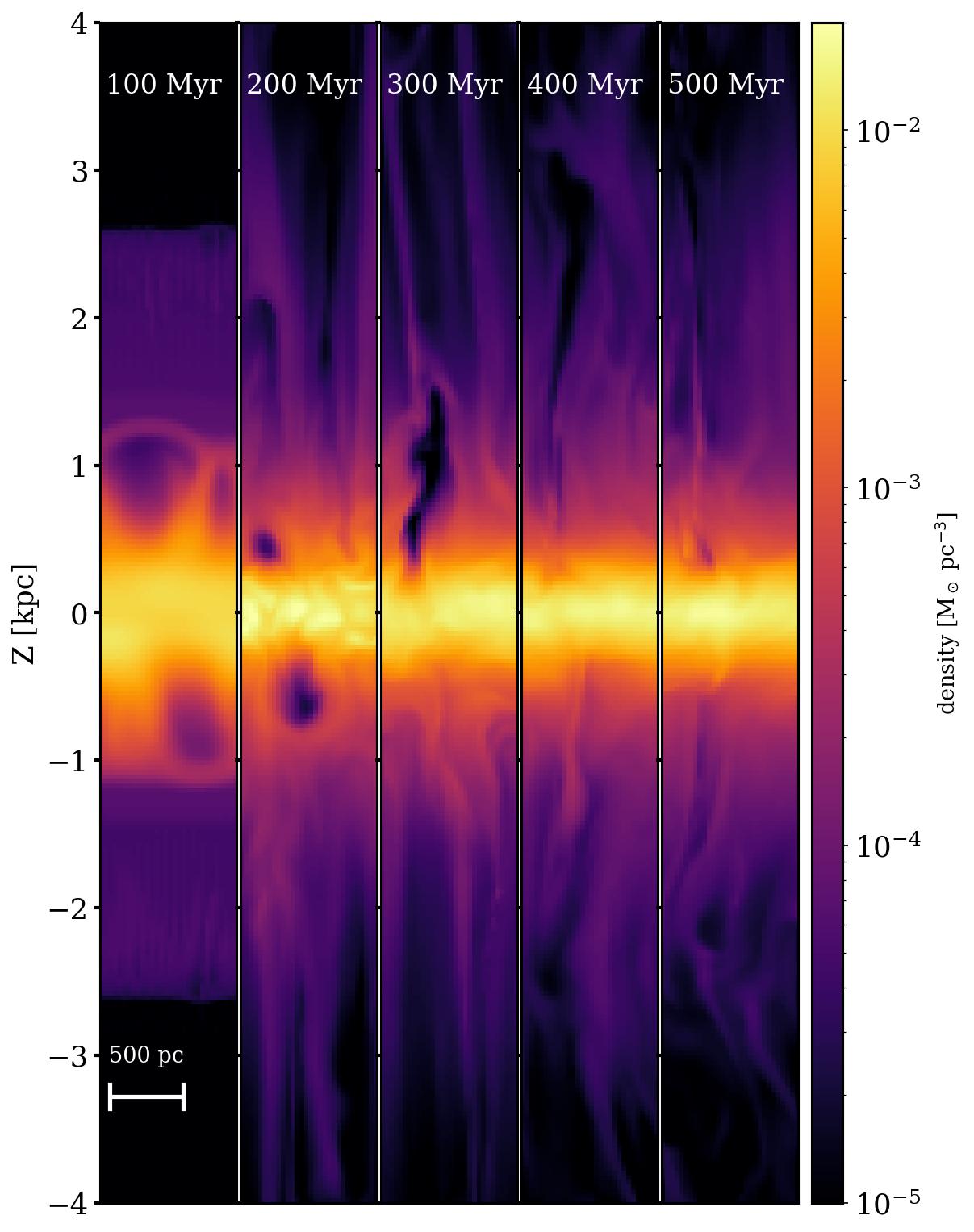
}}
    \centerline{
    \includegraphics[width=.5\linewidth]{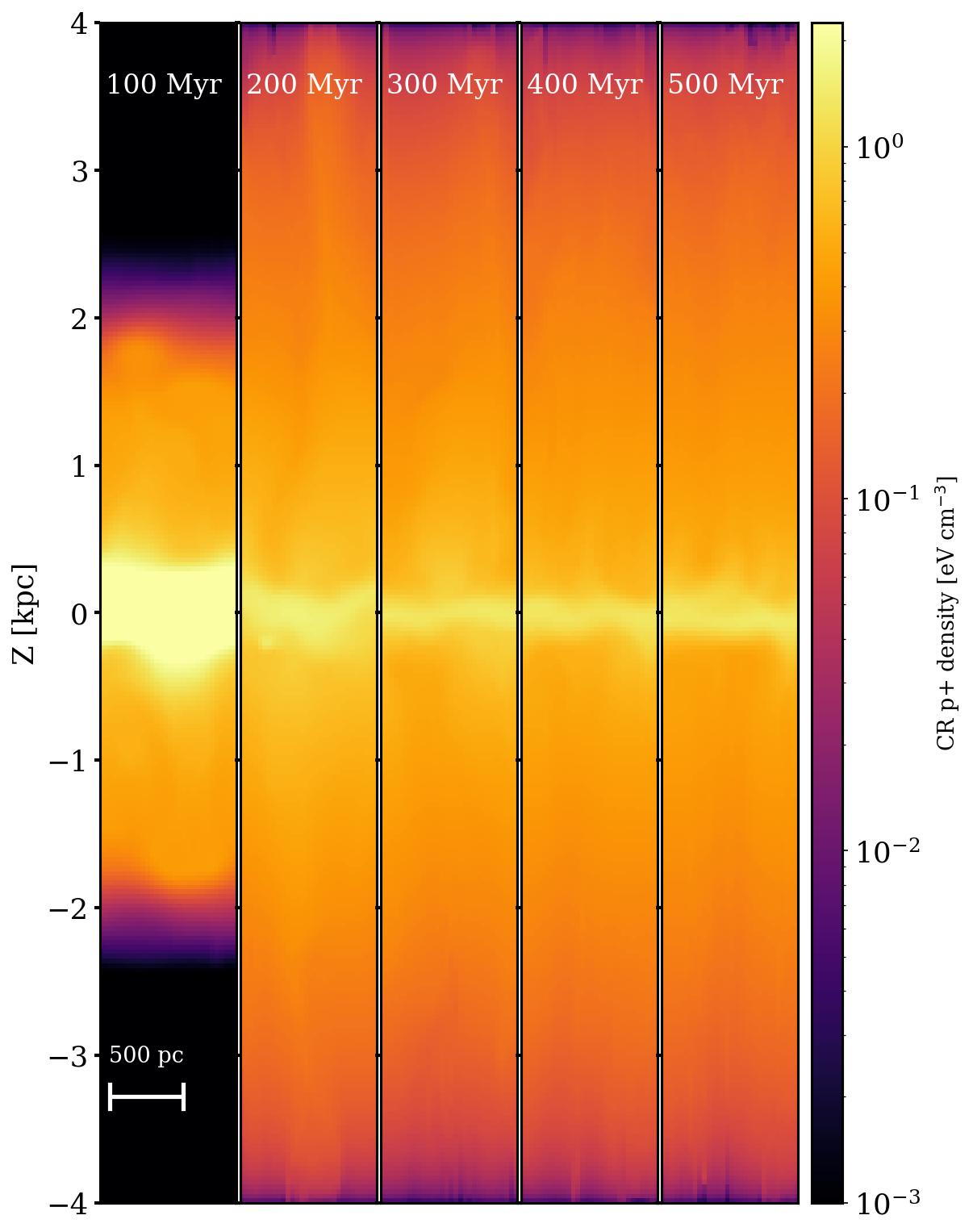
}
    \includegraphics[width=.5\linewidth]{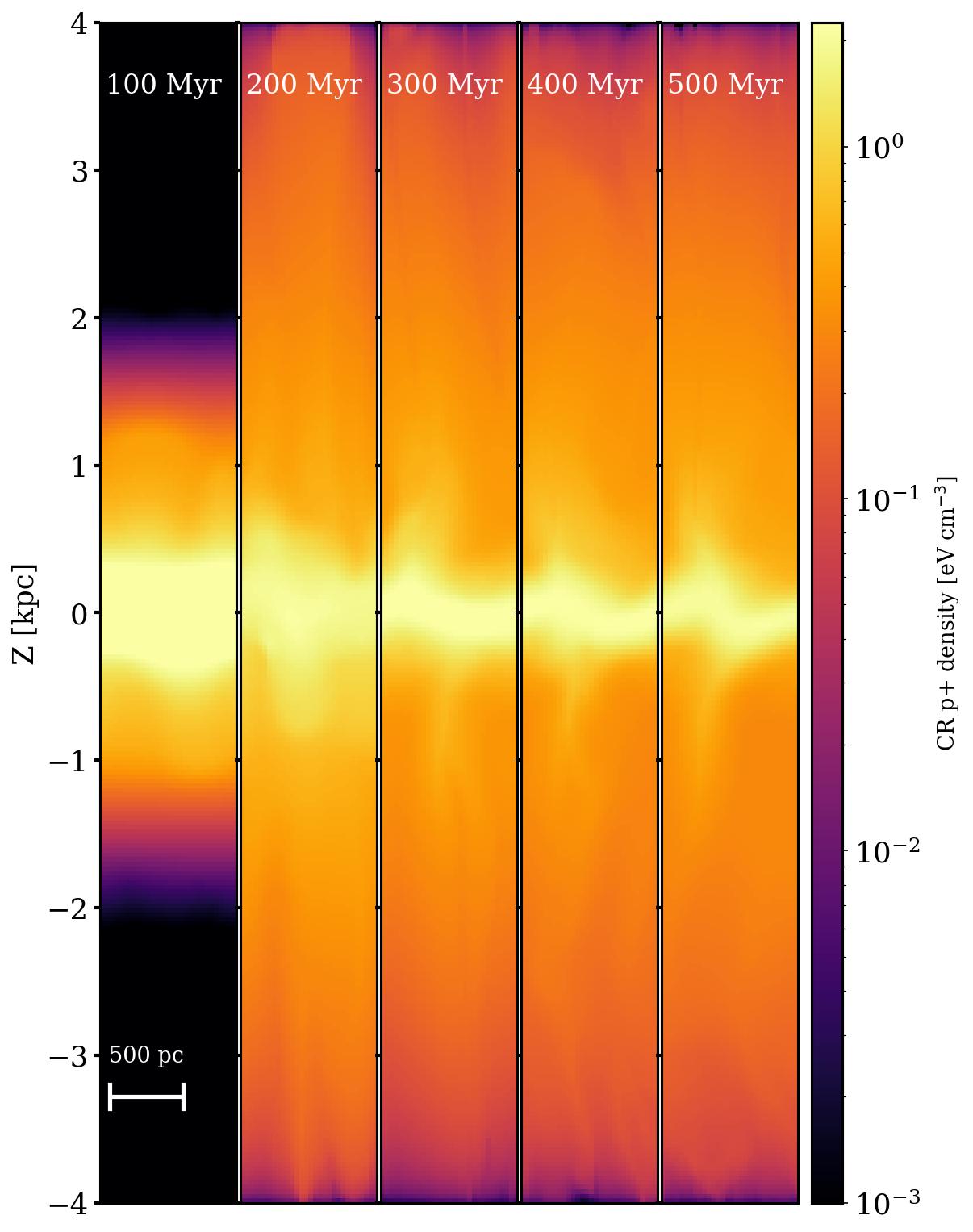
}}
    \caption{Top: snapshots of the gas density evolution in the $y$-$z$ plane ($x=0$) at different epochs for model A (left) with $\eta=0$ and D (right) with $\eta=100\,\pc^2\Myr^{-1}$. Bottom: same as in the top panels, but for CR proton energy density.
    }
    \label{fig.1}
\end{figure*}
\begin{figure*}[ht]
    \centering
    \includegraphics[width=1.\linewidth]{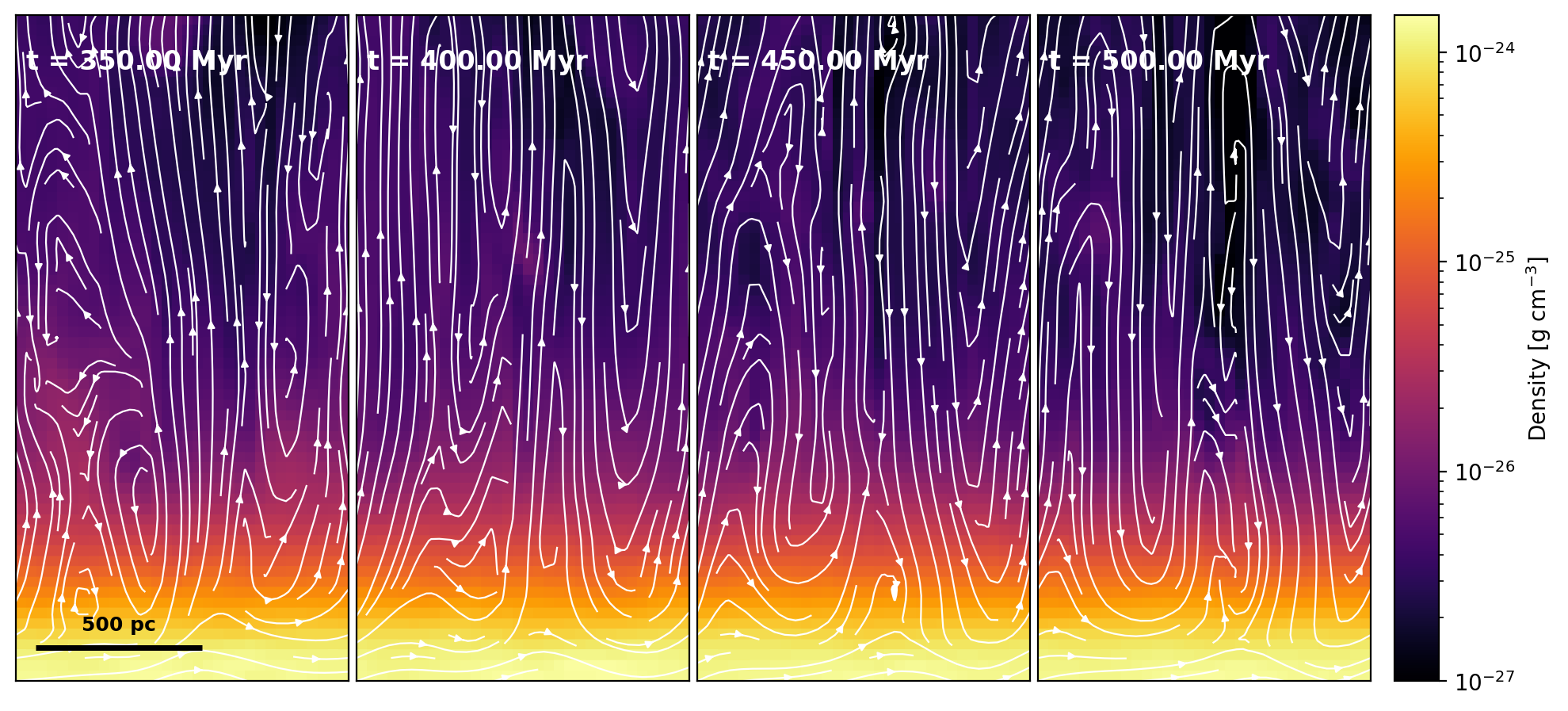
}
    \includegraphics[width=1.\linewidth]{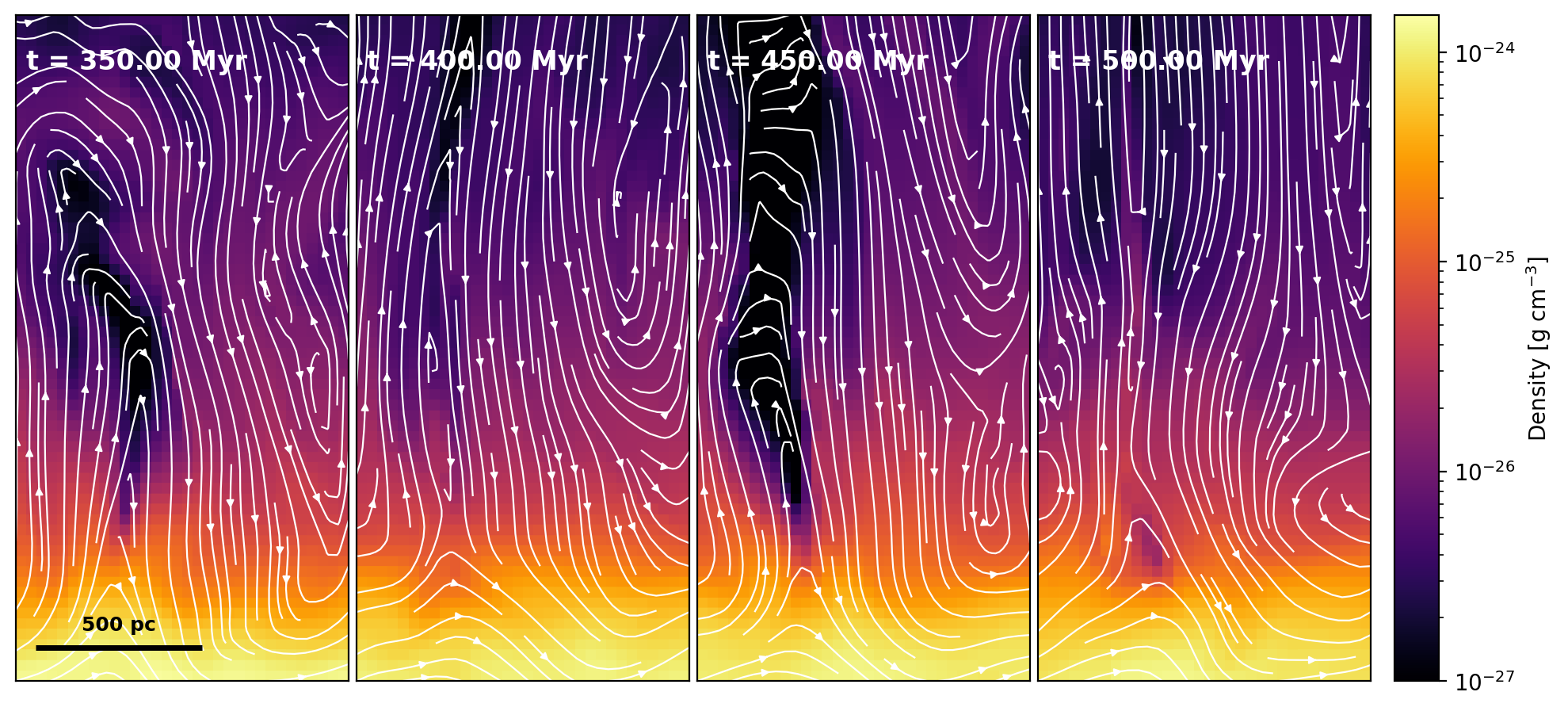
}
    \caption{Snapshots of the gas density evolution within the section $z \in [0,2]\,\mathrm{kpc}$ for models A (upper panels) and D (lower panels), from 350 to 500 Myr, with the magnetic-field lines overplotted.}
    \label{fig.2}
\end{figure*}

We present four simulations with different values for $\eta$. The four tested models are listed in Table \ref{table_1}. The model A follows the ideal-MHD approximation ($\eta = 0$). Fig. \ref{fig.1} displays snapshots of the gas density and CR energy density at $x = 0$ in the $y$--$z$ plane of the stratified box at different times. At $t = 100$ Myr and 200 Myr, both simulations exhibit a strongly perturbed dense disk region due to supernova activity, with buoyant gas structures developing into vertical gas outflows. These outflows are initially driven by the buoyancy of supernova-injected CRs (see discussion in \cite{2026ApJS..284...34B}). After 200 Myr, model A reaches a quasi-steady state in which vertical gas outflows are continuously driven by CRs escaping from the quasi-stationary disc. The CR energy density snapshots support this interpretation: CRs accumulate in the disk at 100 Myr and subsequently escape vertically, as evidenced by the lower CR energy density in the disk and its extension toward higher $|z|$ after 100 Myr. At later times, only a thin disk layer remains visible, persisting until 500 Myr.

In model D, the disk remains strongly perturbed until 500 Myr, in contrast to the ideal-MHD case. Low-density gas structures form after 200 Myr and progressively elongate vertically. The CR energy density snapshots show a dense region centered at $z = 0$, which is denser and thicker in model D, extending to $z \approx \pm 0.5$ kpc, compared with $z \approx \pm 0.25$ kpc in model A. CRs accumulate within the disk and remain confined to it, sustaining the pressure that ejects the gas throughout the simulation.

At 100 Myr, the initially buoyant gas structure has reached a more advanced stage in model A than in model D, while CRs have also propagated to greater altitudes in model A. Since CRs are the only source of feedback in our setup, this suggests that resistivity already affects vertical CR transport in the early stages of model evolution.

To interpret the dynamics in Fig. \ref{fig.1} and explain the less efficient vertical CR transport in resistive MHD, we analyze the magnetic-field structure. Fig. \ref{fig.2} presents the gas density within $z \in [0,2]$ kpc for models A and D from 350 to 500 Myr, at $x = 0$ in the $y$--$z$ plane, with the magnetic field lines shown. For model A, the field lines are predominantly vertical at $z > 0.25$ kpc. In this configuration, CRs above the thin disk layer ($z > 0.25$ kpc) escape vertically. For model D, the magnetic-field structure becomes vertically corrugated in newly formed low-density gas columns (panels at 350 and 450 Myr): the antiparallel vertical fields interact via resistive processes, altering the magnetic topology and creating horizontal bridges between adjacent antiparallel vertical fields. The resulting horizontal magnetic-field structure inhibits the vertical escape of CRs. The low-density structures are consistent with hot gas, heated by magnetic dissipation through Ohmic heating.

\begin{figure}[ht]
    \centering
    \includegraphics[width=1.\linewidth]{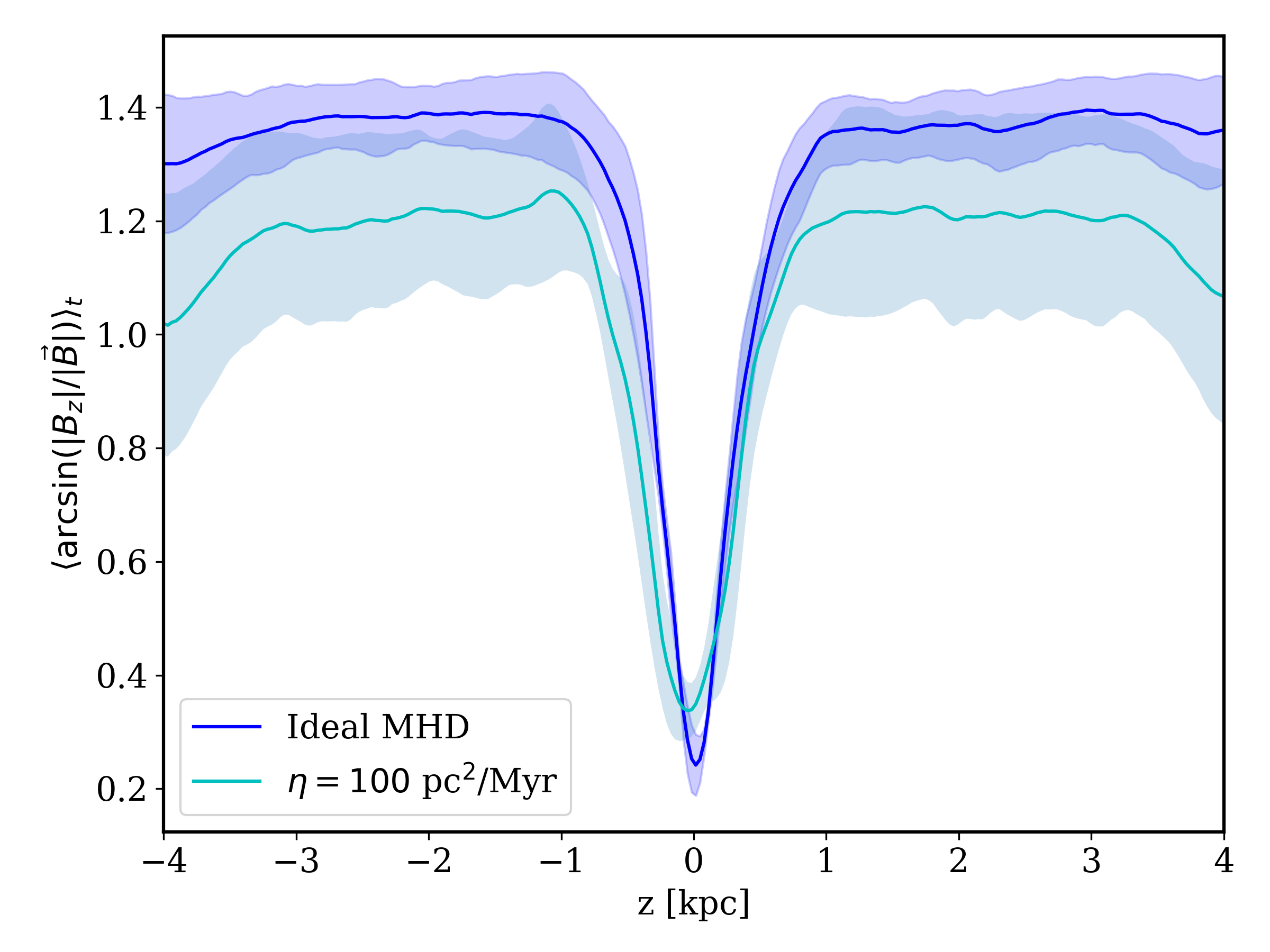}
    \caption{Time-averaged vertical profile of the parameter $\alpha:= \arcsin(|B_z|/|\mathbf{B}|)$ in the $x$--$y$ plane as a function of altitude $z$, displayed with the temporal standard deviation (colored bands). The profile is time-averaged over 210-500 Myr. The data between 0 and 210 Myr are omitted because, during this period, the system transitions from an initially horizontal magnetic field to a vertical magnetic field with alternating orientation in the halo. This unsteady transition is irrelevant and induces substantial variation in the mean and standard deviation.}
    \label{fig.3}
\end{figure}

To strengthen the analysis of the magnetic-field structure, we display for models A and D, in Fig. \ref{fig.3}, the time-averaged vertical profiles of the parameter
\begin{equation}
    \alpha(t) := \arcsin\left(\frac{|B_z|}{|\mathbf{B}|}\right),
\end{equation}
which characterizes the absolute inclination of the magnetic field relative to the horizontal plane. At $|z| > 1$ kpc, the mean value of $\alpha(t)$ is higher in model A than in model D, with $\langle\alpha(t)\rangle \in [1.2,1.4]$ and a standard deviation of $\sigma \in [1.19,1.43]$, compared with $\langle\alpha(t)\rangle \in [1.0,1.21]$ and $\sigma \in [0.8,1.4]$ for model D. This indicates that the magnetic field is, on average, less horizontally oriented in the halo of model A than in model D. Consequently, model D has fewer vertically oriented magnetic-field lines along which CRs can propagate, thereby contributing to the stronger CR confinement observed in Fig. \ref{fig.1}.

Despite this overall trend, Model D nevertheless exhibits substantial temporal fluctuations in $\alpha$. This suggests that although the magnetic field in model D is, on average, more horizontally oriented, its magnetic configuration remains highly time-dependent: as seen in Fig. \ref{fig.2}, in the absence of low-density gas structures for model D (panels at 400 and 500 Myr), the magnetic-field structure relaxes toward a vertically oriented configuration. The magnetic field therefore alternates between predominantly horizontal and vertical configurations.

\subsection{Impact of magnetic reconnection on CR transport}

\begin{figure}[ht]
    \includegraphics[width=1.\linewidth]{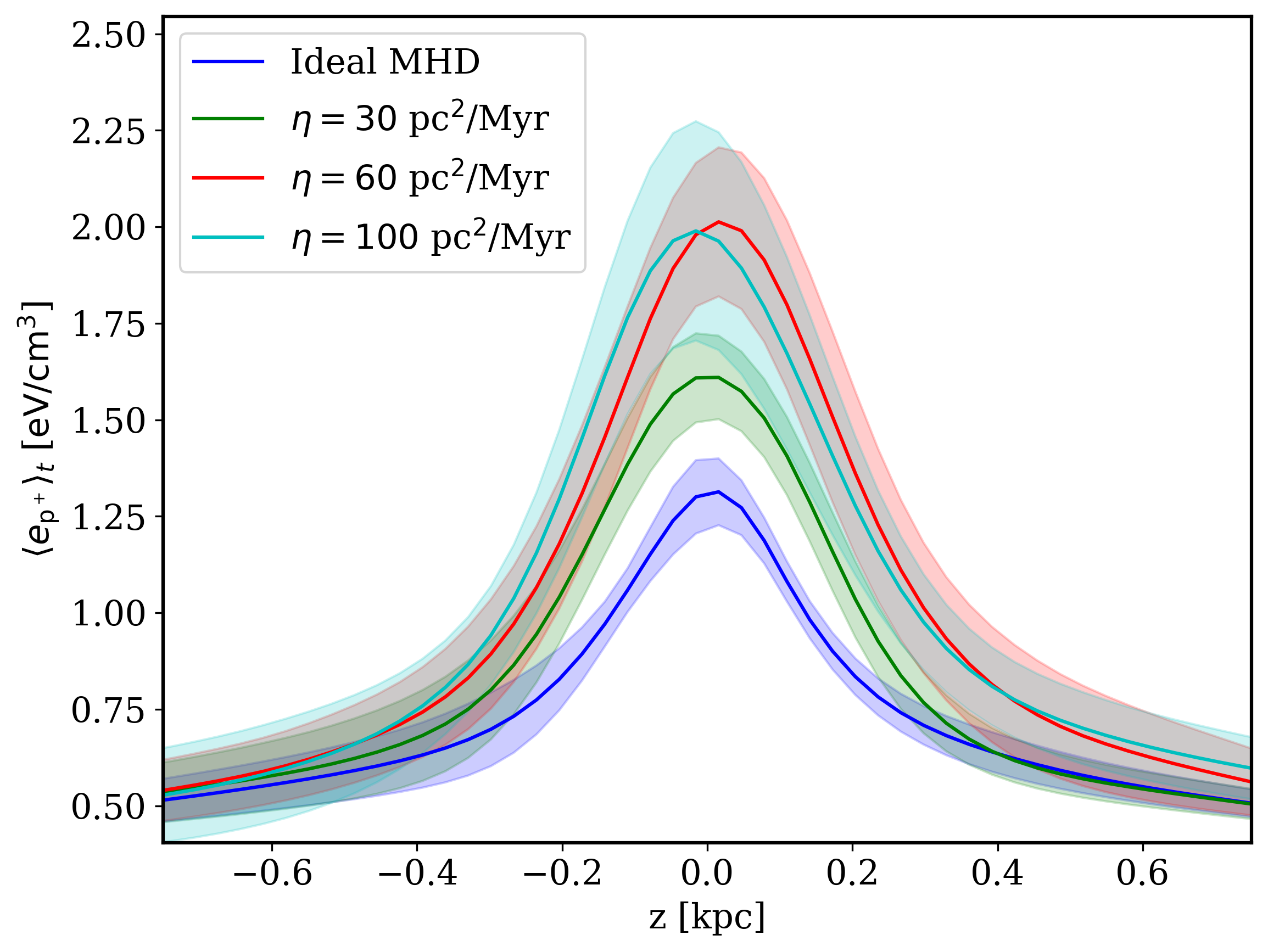
}
    \includegraphics[width=1.\linewidth]{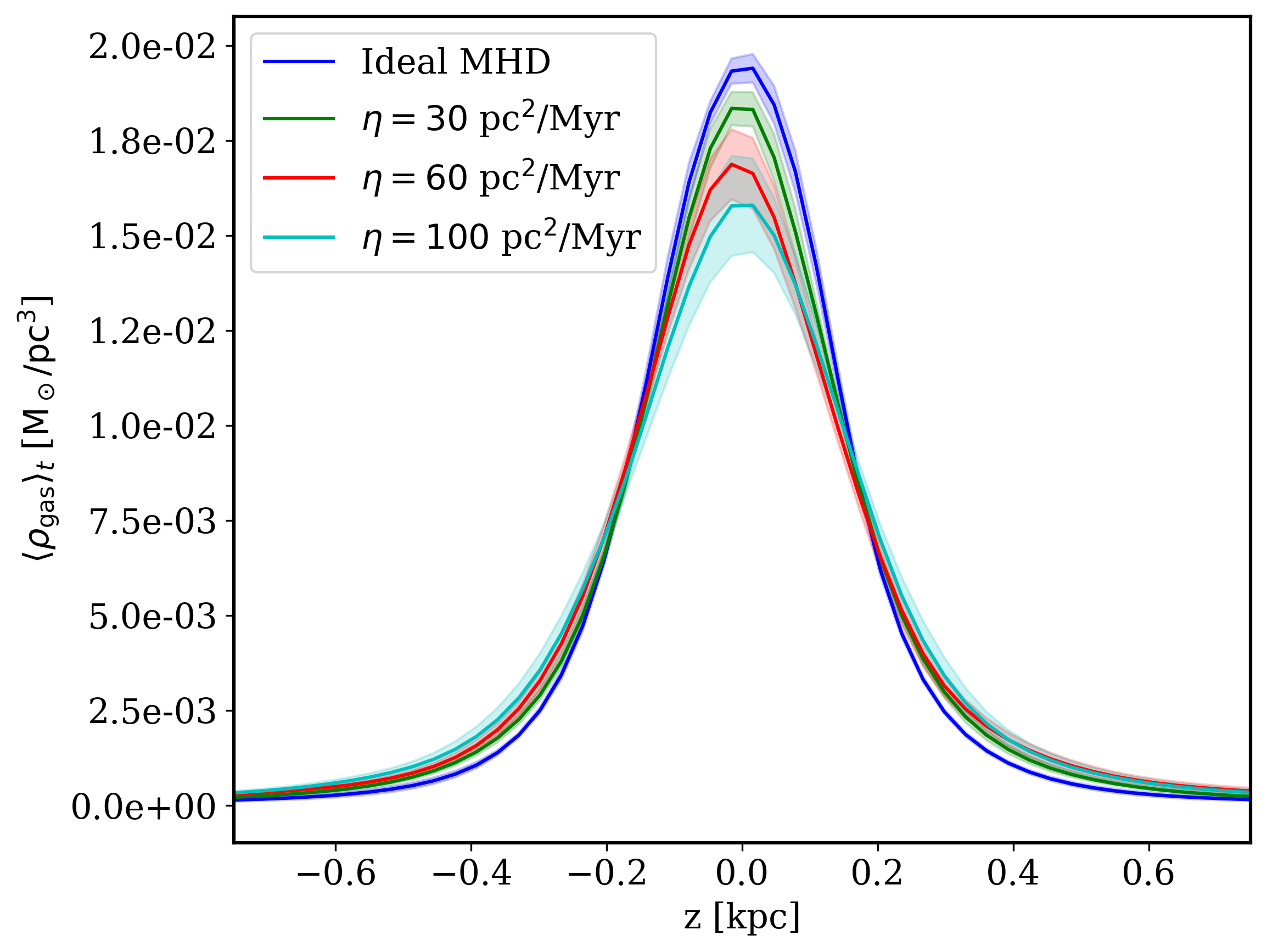
}
    \includegraphics[width=1.\linewidth]{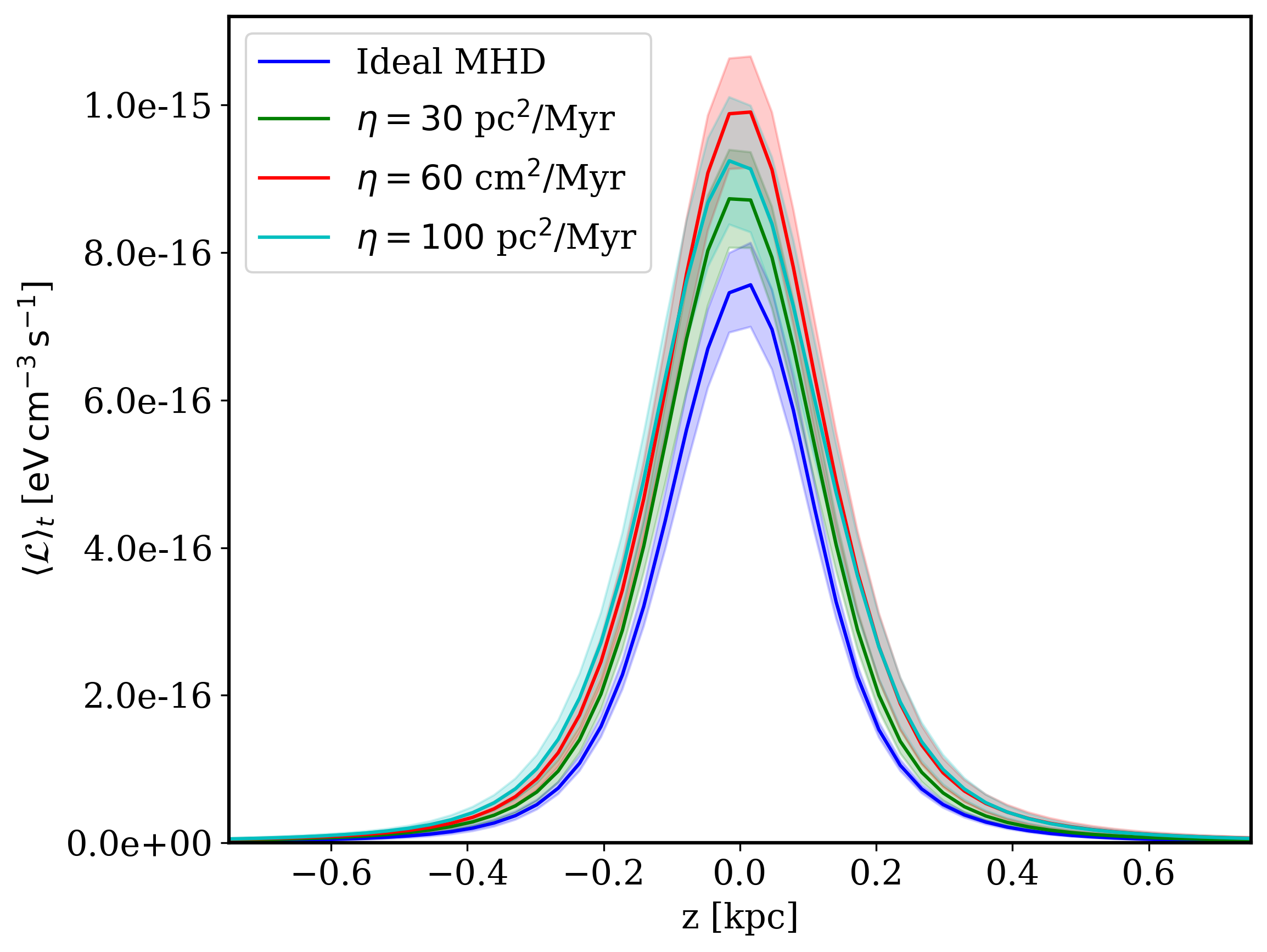
}
    \caption{Time-averaged vertical profiles of the CR proton energy density (upper), gas density (middle), and hadronic luminosity density (lower) in the $x$--$y$ plane as a function of altitude $z$, displayed with the temporal standard deviation (colored bands). The profiles are time-averaged from 210 to 500 Myr.}
    \label{fig.4}
\end{figure}

\begin{figure}[ht]
    \includegraphics[width=1.\linewidth]{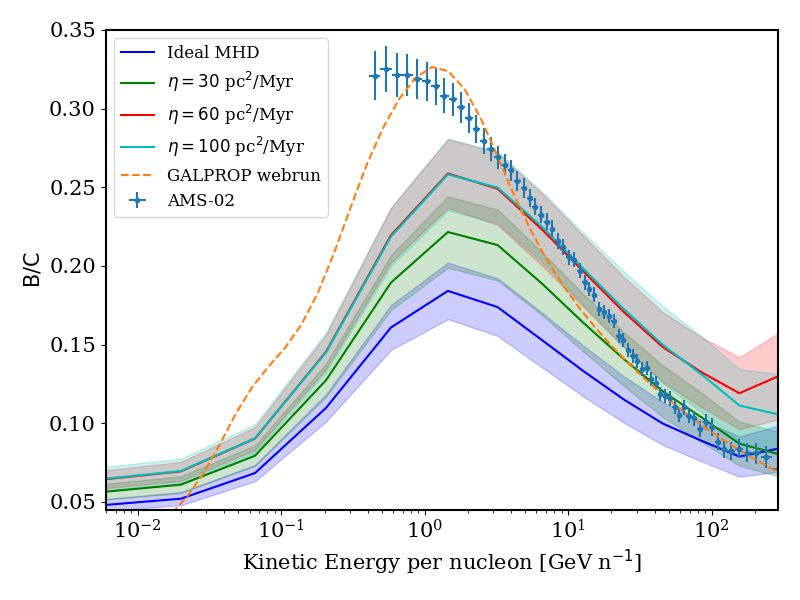
}
    \caption{B/C averaged over the 3D disk region $z \in [-0.3,0.3]\,\mathrm{kpc}$ and time-averaged from 210 to 500 Myr for the four models in Table \ref{table_1}, as a function of kinetic energy per nucleon, compared with the fiducial GALPROP WebRun simulation \citep{2011cophc.182.1156v} and AMS-02 experimental data \citep{2016phrvl.117w1102a}.}
    \label{fig.5}
\end{figure}

Fig. \ref{fig.4} presents the vertical profiles of the CR proton energy density, gas density, and hadronic luminosity density, averaged over time and over the $x$--$y$ plane, for the four models listed in Table \ref{table_1}. The CR energy density is systematically lower in the ideal-MHD system (model A). This behavior is explained by the magnetic structure presented and discussed in the previous section. The more horizontally oriented magnetic field in models B, C, and D inhibits the vertical escape of CRs diffusing along the field lines, leading to their accumulation in the disc. Nevertheless, models C and D exhibit similar energy density distribution, suggesting that the dependency of CR energy density on $\eta$ becomes weaker for the higher tested values of magnetic diffusivity.

The gas density shown in Fig. \ref{fig.4} systematically decreases in the disk as the magnetic diffusivity $\eta$ increases. The dependence of $\rho_{\rm gas}$ on $\eta$ is monotonic across the four models. The lower gas density is consistent with the stronger expansion of trapped
CRs, when reconnection tends to reproduce horizontal fields. The CR confinement
provides larger pressure in the disc, increasing the vertical gas outflows.

For models A and B, the hadronic luminosity density increases with $\eta$. For models C and D, this trend is reversed: model C exhibits a higher luminosity density than model D. Although models C and D have similar CR energy-density distributions, model D has a lower gas density in the central disk due to the gas outflows driven by accumulated CR pressure. The hadronic luminosity density depends on both the gas density and the CR energy density (see Eq. (\ref{eq.3})), so the lower gas density resulting from gas ejection in model D leads to a lower hadronic luminosity density, despite a high CR energy density.

For a better analysis of CR transport, we focus on the
secondary-to-primary ratio B/C. In the leaky-box model \citep{1967PhRv..158.1238C}, B/C depends on the escape time of CRs $\tau_{\rm esc}$, which is related to the momentum-dependent diffusion coefficient, following $\mathrm{B/C}(>1\,\mathrm{GeV\,n^{-1}}) \propto \tau_{\rm esc} = H^2/D(p)$, where $H$ is the vertical height of the considered halo (see details in \cite{2023arXiv230900298E}).

Fig. \ref{fig.5} shows the space-averaged and time-averaged B/C in the disk for the four different models, obtained with diffusion coefficient $D_\parallel^0=3\times10^{28}\,\cm^2\,\s^{-1}$ at nucleus rigidity $R_N=10\,$ GV, and momentum-dependent diffusion power-law index $\delta=0.3$. While those parameters are from the fiducial model in \cite{2026ApJS..284...34B}, we acknowledge that the value of $\delta$ does not coincide with AMS-02 and GALPROP WebRun data; the latter estimates, based on observations, are $\delta=0.5$ \citep{2021PhR...894....1A}. B/C has lower values in the ideal-MHD model (blue curve), with a maximum of $\sim0.17$ at $1\,\mathrm{GeV\,n^{-1}}$. Model B, having $\eta = 30\,\mathrm{pc^2\,Myr^{-1}}$, peaks at $\sim0.19-0.24$, while model C, with $\eta = 60\,\mathrm{pc^2\,Myr^{-1}}$, and model D, with $\eta = 100\,\mathrm{pc^2\,Myr^{-1}}$, peak at $\sim0.24-0.27$. B/C for the latter two models nearly overlap, which exhibits the weaker CR response on $\eta$ observed in Fig. \ref{fig.4}. Despite this, the differences between the models are significant: the B/C mean value and standard deviation across all resistive models are higher than those of model A. This corroborates the results for CR energy density in Fig. \ref{fig.1} and Fig. \ref{fig.4}: in the resistive-MHD models, primary \Ct accumulates in the central region and undergoes increased spallation with the thermal gas, decaying \Ct and producing secondary \Bt\, and \Bel. The secondary abundance relative to the primary population increases, raising B/C.

This analysis demonstrates that the escape time inferred from B/C increases significantly in a turbulent resistive MHD system where magnetic reconnection occurs. Equivalently, the corresponding effective vertical diffusion coefficient decreases.

\section{Discussion and Conclusion}

We investigated the impact of magnetic diffusivity on CR transport using numerical models of gravity-stratified boxes containing ISM gas, magnetic field, and CRs. CR components include spectrally-unresolved protons driving the ISM and undergoing adiabatic, Coulomb, and hadronic energy losses, as well as spectrally resolved \Ct, \Bt, and \Bel nuclei undergoing adiabatic and Coulomb losses. The primary \Ct decays and produces secondary \Bt\, and \Bel through spallation with thermal gas protons. CRs are transported through gas advection and anisotropic diffusion along the magnetic-field lines. To study the effect of magnetic reconnection driven by magnetic turbulence on CR transport, we employed a diffusive magnetic field in the resistive-MHD approximation as a subgrid model of turbulent reconnection. We performed four simulations: one ideal-MHD model and three resistive-MHD models with different values of the magnetic diffusivity $\eta$, motivated by estimates of turbulent magnetic diffusivity in magnetic-arm regions of spiral galaxies \citep{2004astro.ph.11739S, Shukurov_Subramanian_2021}. We then analyzed the CR energy-density profile, hadronic $\gamma$-ray luminosity density profile, and B/C to diagnose the impact of magnetic diffusivity on CR transport.

Our main result is that a change in magnetic topology, triggered by nonzero magnetic diffusivity, enhances CR confinement, leading to higher energy density, hadronic $\gamma$-ray luminosity density, and B/C in the disc. Magnetic reconnection in resistive models regenerates the disk magnetic field's horizontal configuration.
The horizontal field leads to stronger CR accumulation and a longer CR confinement period, and the accumulating CR pressure ejects more gas from the disc. This behavior differs from the ideal-MHD model, which reaches a quasi-stationary state after approximately 200 Myr, with the wind-stretched vertical magnetic field of alternating polarity in the halo.
These results indicate that magnetic dissipation can modify CR transport by affecting the magnetic field topology.

Our findings also corroborate an important conclusion of \cite{2026ApJS..284...34B}. In that study, we showed that the CR escape time inferred from B/C in CR-MHD models does not necessarily reflect the diffusion coefficient used as an input parameter, despite the expectation from the simple relation $\tau_{\rm esc} \propto 1/D$ when $D$ is fixed. In our simulations, $\tau_{\rm esc}$ instead follows the dynamical coupling between CRs, the ISM gas, and the magnetic field. The dependence of B/C on magnetic diffusivity found in the present study provides further evidence for this interpretation. In particular, it shows that physical processes associated with the dynamical evolution and dissipation of the magnetic field, including reconnection, can affect the effective CR transport properties inferred from B/C and $\gamma$-ray measurements. This behavior contrasts with the leaky-box approximation \citep{1967PhRv..158.1238C} and with numerical models that do not include dynamical coupling between CRs and the ISM, such as GALPROP \citep{strong&moskalenko1998}.

These results have further implications for CR transport in spiral galaxies. 
Magnetic reconnection is expected to transform the turbulent magnetic field in the optical arms to a more laminar one in the magnetic arms
 \citep{2007A&A...470..539B, 2019Galax...8....4B}. 
If the CR escape time depends on magnetic reconnection, as suggested by our simulations, then $\tau_{\rm esc}$ may differ significantly between magnetic-arm and inter-arm regions. B/C measurements are obtained in the vicinity of the Solar System, which is located in an inter-arm region. Our results, therefore, suggest that the transport properties inferred from local B/C measurements may primarily characterize the CR transport in this local environment rather than in the Galaxy as a whole. In this picture, $\tau_{\rm esc}$ should be regarded as a spatially dependent property of the CRs rather than as a universal galactic transport parameter.

The simulations also provide a possible observational test of this interpretation. As shown in Fig. \ref{fig.4}, the resistive-MHD models produce higher hadronic $\gamma$-ray luminosity density than in ideal-MHD. Although direct measurements of B/C are limited to the solar neighborhood, $\gamma$-ray emission provides a complementary probe of CRs over larger galactic regions. 
If magnetic reconnection strengthens CR confinement in magnetic arms, the resulting increase in CR number density and hadronic interactions may enhance $\gamma$-ray emission from these regions. Spatially resolved $\gamma$-ray observations of magnetic arms could therefore provide an independent test of the CR confinement inferred from B/C.

Several limitations should be considered when interpreting these results. First, our simulations use gravity-stratified boxes rather than full galactic-disk models. The adopted geometry and boundary conditions influence both CR transport and the ISM's response to magnetic diffusivity. Although the qualitative effect of magnetic diffusivity on CR transport is prominent in our models, a quantitative assessment of its impact on galactic CR transport requires simulations of full, large-scale galactic disk models with realistic magnetic arms and star formation localized in the spiral arms. Second, we model magnetic dissipation at current sheets using a subgrid prescription. Because the numerical resolution is insufficient to resolve the magnetic turbulence directly, the simulations do not explicitly capture the full turbulent reconnection process. Third, we consider only CR-driven gas outflows and exclude thermal or stellar feedback, which may also influence the structure of the ISM and the transport of CRs. Fourth, we do not include spectrally resolved CR protons, which prevents us from calculating the hadronic $\gamma$-ray spectrum that would strengthen the observational diagnostics. Finally, the CR streaming is not included in our models, and we limit CR transport via advection by the thermal gas and anisotropic diffusion in extrinsic magnetic-field turbulence. We plan to address those issues in future work.

\label{subsec:conclusion}

\section{Acknowledgements}

Calculations were carried out at the Center of Informatics—
Tricity Academic Supercomputer \& networK (TASK) and on
the HYDRA cluster at the Institute of Astronomy of Nicolaus
Copernicus University in Torun. A.B acknowledges acknowledges the use of OpenAI’s
ChatGPT to improve the phrasing of portions of the manuscript text, and to assist
with optimization of the post-processing routines. All AI suggestions were
reviewed, validated, and thoroughly tested by the author before adoption. A.B would like to thank Martin Fournier for inspiring discussions.
\section{Data Availability}

The data underlying this paper will be shared on reasonable request to the corresponding author.
\newpage
\appendix

\section{Resistive CR-MHD simulations setup} 
\label{Appendix_A}

Our simulations are carried out with {\scshape Piernik}, a grid-based multi-fluid magneto-hydrodynamical (MHD) code, using conservative numerical schemes\footnote{The public version of {\scshape Piernik} is available at
\href{https://github.com/piernik-dev/piernik}{github.com/piernik-dev/piernik}.}.
{\scshape Piernik} is parallelized based on the Message Passing Interface (MPI) standard, and its data input/output communication utilizes a parallel Hierarchical Data Format 5 (HDF5). The MHD algorithm of {\scshape Piernik} is based on the standard set of resistive MHD equations in conservative form \citep[see eg][]{Shukurov_Subramanian_2021}

\begin{align}
\begin{dcases}
\label{eq.A1}
  & \partial_t \rho + \nabla\cdot\left(\rho\mathbf{v}\right) = 0, \\
  & \partial_t(\rho\mathbf{v}) + \nabla\cdot\left(\rho\mathbf{v}\mathbf{v}^\mathrm{T} - \frac{\mathbf{B}\mathbf{B}^\mathrm{T}}{4\pi}\right) + \nabla P\ = \rho\mathbf{g} - \nabla  P_{\mathrm{p}^+},\\
  & \partial_t e + \nabla\cdot\left[\left(e + P\right)\mathbf{v} - \frac{\mathbf{B}\cdot\mathbf{v}}{4\pi}\mathbf{B}
      + \eta \, \mathbf{J}\times\mathbf{B}\right] = \rho\mathbf{v}\cdot\mathbf{g} - \nabla\cdot\left(P_{\mathrm{p}^+} \mathbf{v}\right), \\
  &\partial_t \mathbf{B} - \nabla \times \left(\mathbf{v}\times\mathbf{B} - \eta \nabla \times \mathbf{B}\right) = 0, \\
  & \nabla \cdot \mathbf{B} = 0,
  \end{dcases}
\end{align}

\noindent where $\rho$ and $n$ are the gas volume and number densities, $\mathbf{v}$ its velocity field, $\mathbf{B}$ the magnetic field, $\mathbf{g}$ the gravitational acceleration, $e= e_\mathrm{k}+ e_\mathrm{t}+ e_\mathrm{m}$ the total energy density,  including kinetic, thermal, and magnetic energy components, $P = P_\mathrm{t}  + P_\mathrm{m}$ is the total pressure, including thermal and magnetic components, and $ P_{\mathrm{p}^+} $ is the CR proton pressure.
The above set of resistive MHD equations is evolved with a finite-volume scheme using the Harten–Lax–van Leer Discontinuities (HLLD) Riemann solver \citep{2005JCoPh.208..315M}, together with the divergence-cleaning algorithm of \citet{2002jcoph.175..645d}, following the implementation described in \citet{2023MNRAS.522.5529P}.
 The MHD energy equation includes the resistive transport of the magnetic energy via the resistive flux $\mathbf{F}_{\rm res} = \eta\,\mathbf{J}\times\mathbf{B}$, in addition to the advective flux. The lack of an explicit magnetic energy dissipation term reflects the presence of Ohmic heating, which converts the magnetic energy into thermal energy. Radiative cooling and heating processes, as well as the viscous stresses, are omitted in the current paper. The gas heating from CR-driven instability \citep{2020ApJ...891..157H} is also neglected.

 The system of MHD equations is solved together with the momentum-integrated CR diffusion-advection equation \cite[see eg][]{2021lrca....7....2h} for the CR particle energy density

\begin{equation}
\label{eq.1}
\partial_t e_{\mathrm{p}^+}  + \nabla \cdot \left( \mathbf{v} e_{\mathrm{p}^+} -  D\nabla e_{\mathrm{p}^+} \right)= \nabla \cdot \mathbf{v}(e_{\mathrm{p}^+} - P_{\mathrm{p}^+}) - \Lambda_\mathrm{cool}e_{\mathrm{p}^+} + Q_{\mathrm{p}^+},
\end{equation}
and the adiabatic equation of state relating the pressure to CR energy density
\begin{equation}
P_{\mathrm{p}^+} = (\gamma_{\mathrm{p}^+} - 1) e_{\mathrm{p}^+}.
\label{eq.2}
\end{equation}
Eq. (\ref{eq.1}) is the transport equation, where $P_{\mathrm{p}^+}$ is the CR pressure, $Q_{\mathrm{p}^+}$ the supernovae source term, $\mathbf{v}$ the gas-velocity field, $D=3\times10^{28}\,\cm^2\s^{-1}$ the CR diffusion coefficient along magnetic-field lines, and $\Lambda_\mathrm{cool} = \Lambda_\mathrm{hadr} + \Lambda_\mathrm{Coul}$ the energy-loss term (hadronic and Coulomb losses). We set $\Lambda_\mathrm{Coul}=1.65\times10^{-16} (n_\mathrm{ISM}/\cm^3)\,\s^{-1}$ \citep{2008MNRAS.384..251G}, and $\Lambda_\mathrm{hadr}=Kn_\mathrm{ISM}\sigma_{pp}^\pi c$, where $K=0.5$ is the collision inelasticity. Eq. (\ref{eq.2}) is the equation of state, where $\gamma_{\mathrm{p}^+}=4/3$ is the relativistic adiabatic index. The simulations presented in this paper include gas, magnetic fields, spectrally unresolved protons, and spectrally resolved stable carbon and boron nuclei. The primary and secondary nuclear components are treated using a system of equations based on the Fokker-Planck equation and the numerical methods outlined in \cite{2026ApJS..284...34B}.

\section{Numerical scheme for the diffusive magnetic field in resistive MHD}
\label{Appendix_B}
In this appendix, we detail the numerical scheme used to solve the resistive-MHD system given by Eq. (\ref{eq.A1}).
We advance these equations in a Strang-split manner \citep{doi:10.1137/0705041}, $\mathcal{R}(\Delta t/2)\mathcal{H}(\Delta t)\mathcal{R}(\Delta t/2)$ where $\mathcal{R}$ is the resistive half-time source step of the induction equation, followed by the full-time ideal MHD transport step $\mathcal{H}$, then another source step. The MHD energy equation is
modified within $\mathcal{H}$ to account for the Ohmic heating of the gas by adding the resistive analog of the Poynting flux,
$\mathbf{F}_{\rm res} = \eta\,\mathbf{J}\times\mathbf{B}$ to the total energy flux. The explicit diffusive stability limit additionally constrains the global timestep, $\Delta t \lesssim C_\eta\, \Delta x^2/(2\eta_{\max})$, in addition to the usual advective CFL condition. We set $C_\eta$ = 0.4 in all our runs.

Below, we present two validation test runs of the above scheme. We compare each numerical test with a known analytical solution. We set $\eta = 0.05$ for both runs to avoid large diffusion of the magnetic field that can cause boundary effects. All quantities are mentioned in code units.

\subsection{Diffusion of a 1D Gaussian Magnetic field pulse}
In this test, we consider a static ISM gas ($\mathbf{v}$ = 0) with uniform normalized mass density $\rho = 1.0$ and pressure $P = 1/ (\gamma -1)$ in a periodic 1D domain with 256 unit cells. With $B_x = B_z = 0$ and $B_y = B_y(x)$, the induction equation in Eq. (\ref{eq.A1})
reduces, for uniform $\eta$, to the scalar diffusion equation
$\partial_t B_y = \eta\, \partial_x^2 B_y$. For the Gaussian initial
condition $B_y(x,0) = b_0 \exp(-a x^2)$, the solution is
\begin{equation}
	B_y(x,t) = \frac{b_0}{\sqrt{1+4a\eta t}}
	\exp\!\left(-\frac{a x^2}{1+4a\eta t}\right).
	\label{eq:gaussian_diffusion_solution}
\end{equation}
\begin{figure}[h]
	\centering
	\includegraphics[width=.8\textwidth]{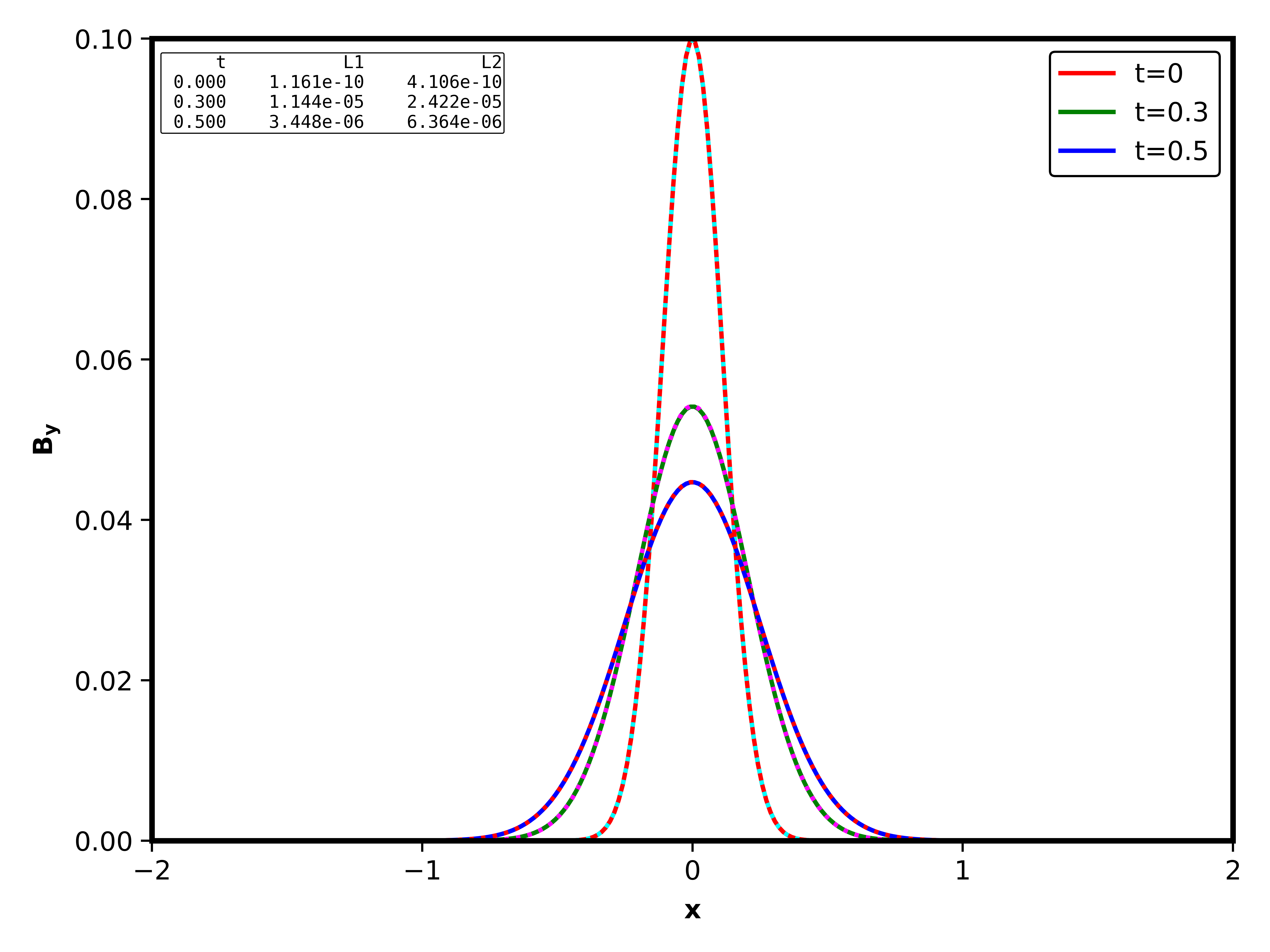}
	\caption{1D Gaussian pulse diffusion: numerical (solid) vs.\
		analytic (dotted) $B_y(x,t)$ at $t=0,\,0.3,\,0.5$, with $L_1$/$L_2$.}
	\label{fig:1d_diffusion}
\end{figure}
In this test, we set $a = 40$ and $b_0 = 0.1 $. Fig.~\ref{fig:1d_diffusion} shows $\mathbf{B_y}(x,t)$ at t = 0, 0.3 and 0.5. We see that it agrees well with the closed-form solution given by Eq. (\ref{eq:gaussian_diffusion_solution}). To quantify the agreement, we compute the $L_1$ and $L_2$ error norms over the
$N$ cells of the uniform grid,
\begin{equation}
	L_1 = \frac{1}{N}\sum_{i=1}^{N} \left| B_{y,i} - B_y^{\rm exact}(x_i,t) \right|,
	\qquad
	L_2 = \left[ \frac{1}{N}\sum_{i=1}^{N} \left( B_{y,i} - B_y^{\rm exact}(x_i,t) \right)^2 \right]^{1/2},
	\label{eq:error_norms}
\end{equation}
where $B_{y,i}$ is the numerical value in cell $i$ and $B_y^{\rm exact}$ is the
analytic solution evaluated at the cell center $x_i$. We see that both are $\mathcal{O}(10^{-6})$ in code units.

\subsection{Diffusion of a 2D Sinusoidal Magnetic Field}
We consider a static ($\mathbf{v}$ = 0), uniform background ($\rho = 1.0, P = 1/ (\gamma -1)$). The initial magnetic field vector is configured as:
\begin{eqnarray}
	B_x(x, y, 0) &=&  B_0\, k_y \sin(k_x x) \cos(k_y y), \\
	B_y(x, y, 0) &=& -B_0\, k_x \cos(k_x x) \sin(k_y y).
\end{eqnarray}

with $B_0 = 0.01$ and $k_x = k_y = 2\pi$. This configuration is explicitly divergence-free. Under uniform resistivity $\eta$, the linear induction equation reduces to decoupled 2D diffusion equations, yielding the exact time-dependent decay:
\begin{equation}
	\mathbf{B}(x, y, t) = \mathbf{B}(x, y, 0) \exp\!\left(-\eta (k_x^2 + k_y^2) t\right).
	\label{eq:2d_sine_solution}
\end{equation}

Consequently, the volume-averaged magnetic energy density decays exponentially according to $\langle B^2 \rangle / 2 \, (t) = E_{mag}(0) \exp\!\left(-2\eta (k_x^2 + k_y^2) t\right)$, where $E_{mag}(0)$ is the initial magnetic energy density.
\begin{figure}[h]
	\centering
	\includegraphics[width=1.\textwidth]{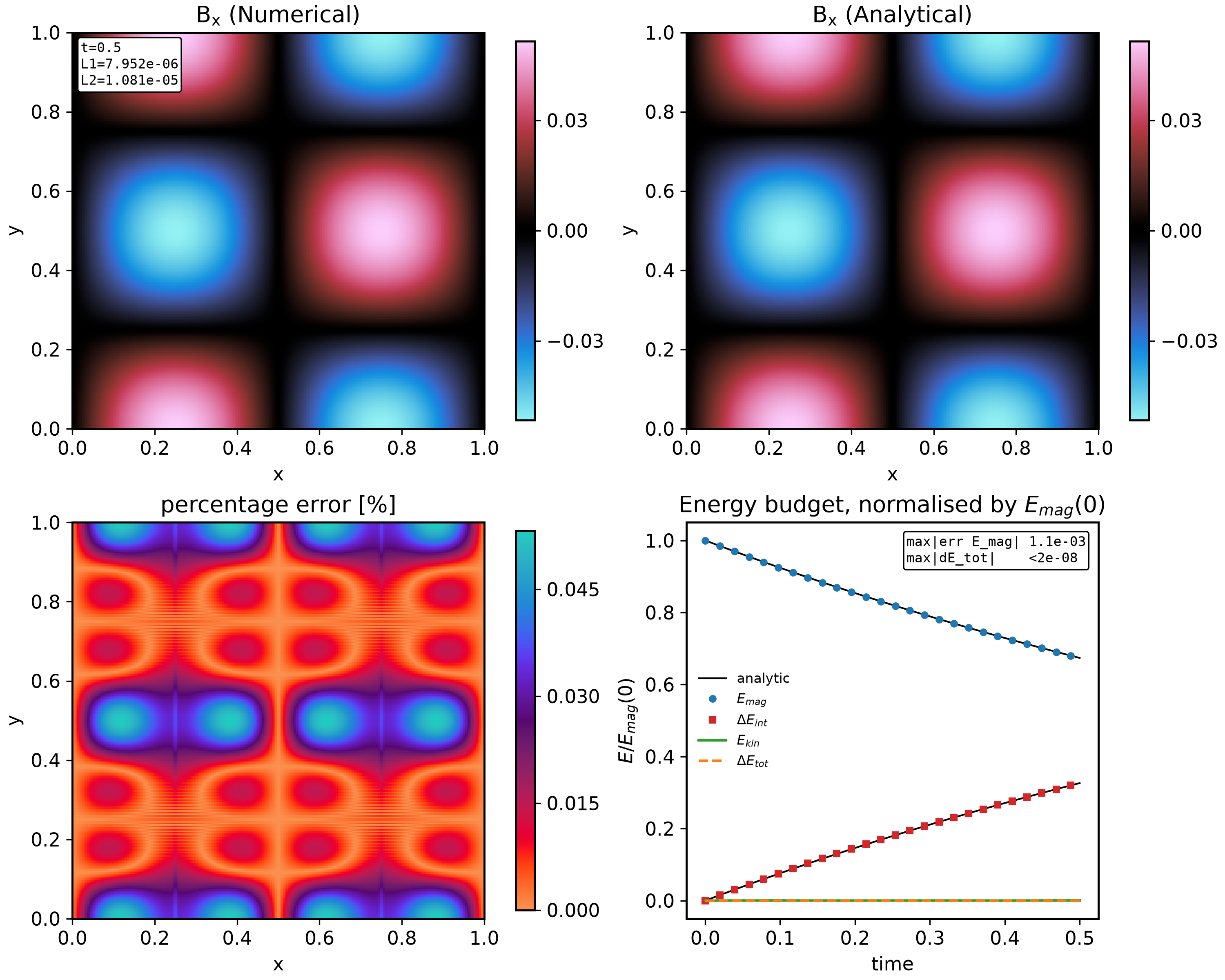}
	\caption{2D Sinusoidal field Diffusion: (top left) $\mathbf{B_x}$ (Numerical) at t = 0.5, (top right) $\mathbf{B_x}$ (Analytical) at t = 0.5, (bottom left) Percentage error = $100 \times |(\mathbf{B_x}^{num} -\mathbf{B_x}^{ana})/\mathbf{B_x}^{ana} |$, (bottom right) decay of volume-averaged magnetic energy $\langle B^2 \rangle / 2$ over time compared against the analytical prediction.}
	\label{fig:2d_diffusion}
\end{figure}
Fig.~\ref{fig:2d_diffusion} shows the comparison between numerical and analytical
results. The numerical $B_x$ agrees well with the analytical profile across the
domain, with L1/L2 errors $\sim \mathcal{O}(10^{-5})$ and a peak percentage error
of $\approx 0.05\%$. The magnetic energy follows the theoretical dissipation rate
to within $\sim 10^{-3}\,E_{mag}(0)$. The dissipated magnetic energy is fully
converted to internal energy by Ohmic heating, and the kinetic energy stays
negligible ($E_{kin} \lesssim 3\times10^{-4}\,E_{mag}(0)$). The total energy is
conserved to the output precision of the log, $|\Delta E_{tot}| < 2\times10^{-8}\,E_{mag}(0)$.

\newpage
\bibliography{Paper}{}
\bibliographystyle{aasjournalv7}

\end{document}